\documentclass[pra,11pt,onecolumn,superscriptaddress,showpacs]{revtex4-1}
\usepackage{amsmath,multirow,amssymb,graphicx,graphics}
\usepackage{times}

\begin{document}

\title{Interferometric detection of Chern numbers in topological optical lattices}

\author{Jian Xu}
\affiliation{Science School, Guangdong Ocean University, Zhanjiang 524088, China}
\affiliation{Guangdong Provincial Key Laboratory of Quantum Engineering and Quantum Materials and School of Physics and Telecommunication Engineering, South China Normal University, Guangzhou 510006, China}


\date{\today}

\begin{abstract}
Topological states of matter emergent as a new type of quantum phases, which can be distinguished by their associated topological invariants, e.g.,  Chern numbers. Currently, there has increasing interests toward the physically detection of the new predicted topological phases. Here, we propose an interferometric approach to directly measure the Chern number in a topological optical lattice via detecting the associated Zak phases. We show that this interferometric approach can distinguish  Zak phases of $\pm2\pi$ from 0 in the first Brillouin zone, and thus provides a new tool to directly detect the Chern number of topological systems. In addition, we demonstrate that this method is feasible under realistic experimental conditions and may  generalize to detect topological systems with higher Chern numbers.
\end{abstract}


\maketitle


\section{Introduction}

Topology of quantum systems is a key concept in modern quantum many-body physics. For example, quantum Hall effect has attracted much attention since it was discovered \cite{K. v. Klitzing,D. C. Tsui}, where the Hall conductance
is proportional to the Chern number (CN) $C$, defined as the integral of Berry curvature over the first Brillouin zone (BZ) \cite{J. E. Avron,B. Simon,R. B. Laughlin,D. J. Thouless}. Specially, Haldane showed that quantum Hall effect can exist without the external magnetic field and the associated Landau levels \cite{F. D. M. Haldane}. Meanwhile, it has also find important application in topological quantum computation \cite{tqc1,tqc2,tqc3,tqc4,tqc5}. On the other hand, the system with cold atoms in optical lattices is a powerful tool to simulate strongly correlated many-body models \cite{mb1}. Recently, synthetic gauge fields and spin-orbit coupling have been experimentally realized in cold atomic gases  \cite{mb2}. These new developments allow one to study many exotic topological phases in condensed matter physics using cold neutral atoms trapped in optical lattices. As remarkable examples, the implementation of the spin Hall effect are presented in cold atoms \cite{S. L. Zhu,she1,she2}, the Haldane model in cold atom system has been theoretically proposed \cite{L.B.Shao,h1} and experimentally realized recently \cite{G. Jotzu}.

The topological phases of many-body systems can be characterized by the CN, which can be manifested with the Hall conductance. Due to the absence of local orders, topological phases are typically hard to detect. Unlike condensed matter systems where a routine measurement of the Hall conductance reveals the CN,  the countable conduction channels has not be realized in cold atoms and the quantized Hall conductance is extremely difficult to be observed \cite{J.-P. Brantut}. Therefore, nowadays, measuring the Hall conductance is technically unrealistic for cold atoms in optical lattices. On the other hand, alternative strategies for the detecting the CN include measuring the time evolution of the center of mass \cite{A. Dauphin}, the gapless edge modes \cite{N. Goldman1,M. Buchhold,N. Goldman2}, the pumped charge \cite{L. Wang}, the Landau-Zener-St\"{u}ckelberg tunneling \cite{L.-K. Lim} and the bulk Chern number from Berry's curvature over the BZ \cite{E. Alba,H. M. Price,D. A. Abanin}. However, these methods either rely on the weak experimental signals or depend on complicated manipulations/measurements on the whole bulk band. For example, in Ref. \cite{D. A. Abanin}, the BZ is divided into many small areas, and the Berry¡¯s curvature of each area is measured separately. But, near Dirac points, the nonadiabatic transition due to the small energy gap leads to the experimental difficulty. Therefore, it is still challenging for the delicate cold atom systems \cite{X.-J. Liu1}. Nevertheless, Ref. \cite{X.-J. Liu1} demonstrated that it can be inferred by measuring of the energy spectrum at highly symmetric points of the BZ. Furthermore, it can be demonstrated that the system is topologically nontrivial when $\nu=1$, but it can not distinguish the case of CN=0 (topologically trivial) and CN=2 (topologically nontrivial) when $\nu=0$. Although one can not build a cold atom system with strict Hall conductance, a similar scheme detecting the CN by measuring the anomalous velocity has been experimentally realized \cite{M. Aidelsburger}.

It is well-known that the CN is not only relate to the Hall conductance but also the Zak phase of the BZ, which is the Berry's phase picked up by a particle moving across the first BZ \cite{J. Zak1}. Here, according to the definition of the CN, we introduce a scheme to detect the CN in topological optical lattices through measuring the Zak phases. Although our scheme shares the same interferometric method as same as that of in Ref. \cite{D. A. Abanin},  we propose to use the noncyclic trajectories of cold atoms from the full controllable moving of cold atoms over the BZ \cite{L. Duca}, which can be realized by a combination of the gradient magnetic field \cite{L. Tarruell} and the accelerating optical lattice \cite{K. W. Madison}. Base on the associated noncyclic Zak phases \cite{J. Zak2}, we propose a interferometric scheme to detect the $\pm \pi$ phase when the trajectories inclose the Dirac points through measuring the geometric phase of a half of Dirac point. In our scheme, unlike previous ones \cite{D. A. Abanin}, we can  distinguish the $\pm \pi$ phases and avoid the problem of nonadiabatic transitions. In particularly, based on this distinguishing, it  provides a method to directly detect the CN in cold atom systems. In addition, our scheme can also realize complex operation, such as one spin state is pinned and the other spin state is moved to obtain Zak phase. Therefore, it provides a newly possible idea to detect the  $\mathbb{Z}_2$ topological invariants \cite{J. de Lisle,F. Grusdt,S.-T. Wang}.

\section{The detection of topological properties}

\subsection{The Haldane model and its CN}
We consider a conventional topological model with a honeycomb lattice. For bosons in optical lattices, considering the simplest case, the Hamiltonian of the model is described by
\begin{equation}
\label{Haldane model}
H=H_0(k)I+H_x(k)\sigma_x+H_y(k)\sigma_y+H_z(k)\sigma_z,
\end{equation}
with
\begin{subequations}
\begin{equation}
H_0=-2t'\cos\phi[\cos(\mathbf{k}\cdot \overrightarrow{v_1}+\cos(\mathbf{k}\cdot \overrightarrow{v_2})+\cos(\mathbf{k}\cdot \overrightarrow{v_3})],
\end{equation}
\begin{equation}
H_z=-2t'\sin\phi[\sin(\mathbf{k}\cdot \overrightarrow{v_1})+\sin(\mathbf{k}\cdot \overrightarrow{v_2})+\sin(\mathbf{k}\cdot \overrightarrow{v_3})],
\end{equation}
\begin{equation}
H_x=-t[\cos(\mathbf{k}\cdot \overrightarrow{e_1})+\cos(\mathbf{k}\cdot \overrightarrow{e_2})+\cos(\mathbf{k}\cdot \overrightarrow{e_3})],
\end{equation}
\begin{equation}
H_y=-t[\sin(\mathbf{k}\cdot \overrightarrow{e_1})+\sin(\mathbf{k}\cdot \overrightarrow{e_2})+\sin(\mathbf{k}\cdot \overrightarrow{e_3})],
\end{equation}
\end{subequations}
where $\overrightarrow{e_1}=(0,a)$, $\overrightarrow{e_2}=(-\sqrt{3}a/2,-a/2)$ and $\overrightarrow{e_3}=(\sqrt{3}a/2,-a/2)$ are the lattice vectors, $\overrightarrow{v_1}=(\sqrt{3}a,0)$, $\overrightarrow{v_2}=(-\sqrt{3}a/2,3a/2)$ and $\overrightarrow{v_3}=(-\sqrt{3}a/2,-3a/2)$ are the vectors of next-nearest-neighbor hopping, $a$ is the basis vector of lattices, $t$, $t'$ and $\phi$ are adjustable parameters. It is well known that the reciprocal lattice vectors of honeycomb lattice are $\overrightarrow{b_1}=(0,4\pi/(3a))$ and $\overrightarrow{b_2}=(2\pi/(\sqrt{3}a),-2\pi/(3a))$. There are two types of Dirac points $K$ and $K'$ where $K(K')=(\pm4\pi/(3\sqrt{3}a),0)$ in this model. Each Dirac point has $\pm\pi$ Berry's phase,  the sign of which, negative or positive, is determined by the system's parameters. The definition of the CN is $C=\frac{1}{2\pi}\int_{BZ} \Omega d\emph{s}$ with $\Omega=i \nabla\times \langle u(k)|\partial_k|u(k)\rangle$ being the Berry curvature. For the Haldane model, we find that the CN is
\begin{align}
C=\left\{
\begin{tabular}{cc}
1 &  for $\phi \in (0, 2\pi)$, \\
0 & for $\phi=0$, \\
-1 & for $\phi \in (-2\pi, 0).$
\end{tabular}
\right.
\end{align}

\begin{figure}[tbp]
\includegraphics[width=15cm]{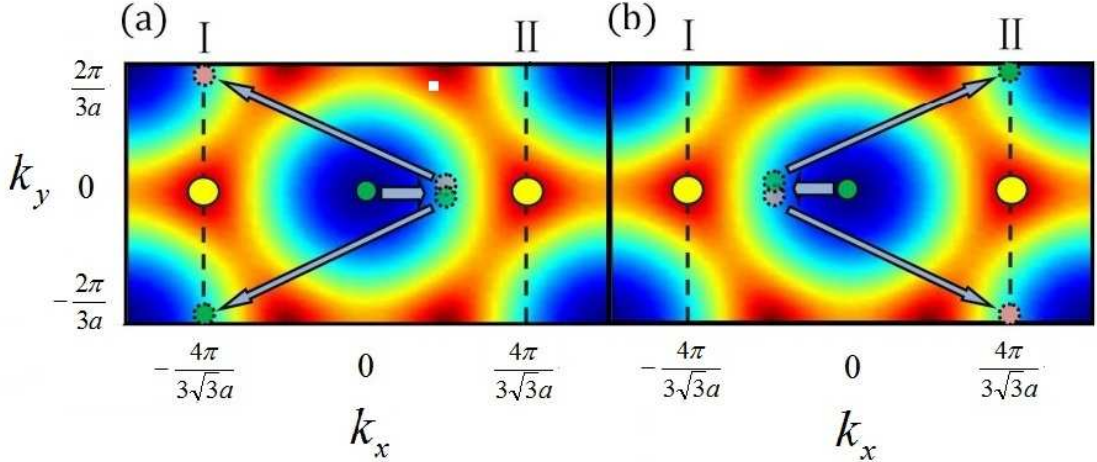}
\label{fig1}
\caption{(Color online) Sequences for detecting the CN in the Hadane model. Suppose a cloud of cold bosonic atoms in a optical lattice is initially prepared in the Hadane phase and the bosonic gas of the spin state $|\downarrow\rangle$ is in the bottom of the energy band. (a) Atoms are firstly moved to the site at $k_x=2\pi/(3\sqrt{3}a), k_y=0$ by a force F induced from the applied gradient magnetic field. Then a $\pi/2$ pulse creates a coherent superposition state of the atomic spin states. After that, the atoms are evolve  to the sites of ($k_{x,(\downarrow,\uparrow)}=-4\pi/(3\sqrt{3}a), k_{y,(\downarrow,\uparrow)}=\pm2\pi/(3a)$) due to the spin-dependent force. Finally, a $\pi/2$ pulse with a phase $\phi_{MW}$ is applied and then the noncyclic Zak phase of the superposition state $\phi_{Zak}^{\mathrm{\uppercase\expandafter{\romannumeral1}}}$ is read out. (b) Atoms are now moved to the site at ($k_x=-2\pi/(3\sqrt{3}a), k_y=0$). After the $\pi/2$ pulse and the combination of the two different forces, the superposition state evolves to the sites of ($k_{x,(\downarrow,\uparrow)}=4\pi/(3\sqrt{3}a), k_{y,(\downarrow,\uparrow)}=\mp2\pi/(3a)$). The noncyclic Zak phase $\phi_{Zak}^{\mathrm{\uppercase\expandafter{\romannumeral2}}}$ can then be read out.}
\end{figure}

For the Haldane model, as the Berry curvatures only distribute within the area of the Dirac cones due to the power law distribution form \cite{D. Xiao}, so an area that is bigger than that of the Dirac cone, which is generally tens of that of the first BZ, is enough to achieve an approximate integer quantization geometric phase. Thus, the definition of the CN can be approximated   by $C=\frac{1}{2\pi}\int_{S} \Omega d\emph{s}$, where $S$ is an area including all Dirac cones within the first BZ. We next link the CN to the geometric phase, i.e., the Zak phase \cite{J. Zak1}, of a open trajectory in the first BZ. In this odel, due to the fact that the geometric phase of the corresponding CN is $2\pi C$, the interferometry can not distinguish this phase in the case of a closed trajectory $S$. However, if we divided the closed trajectory $S$ into several gauge-invariant open trajectories, the total geometric phase of the closed trajectory can be obtained by adding those of open trajectories together. In addition, geometric phase of an open trajectory can be measurable by using an atomic interferometry. In our scheme, for a given wavefunction $u(k)$ with quasimomentum $k$, as shown in Fig. 1, the evolution trajectories and the straight trajectory between the end points $k_b$ and $k_e$ of the trajectories are the reciprocal lattice vector $\pm \overrightarrow{b_1}=(0,\pm 4\pi/(3a))$. The noncyclic Zak phase of an arbitrary trajectory $\phi_{Zak}(\subset)$ can be approximated as \cite{J. Zak2}
\begin{equation}
\label{zak2}
\phi_{Zak}(\subset)=\oint_S\Omega d\emph{s}-\int_{k_b}^{k_e}\langle u(\emph{k})|\partial_k|u(\emph{k})\rangle d\emph{s},
\end{equation}
where $\oint_S d\emph{s}$  is the integral over the closed trajectory, constituted by $u(k)$ and the straight trajectory connecting  $k_b$ and $k_e$ points, and $\int_{k_b}^{k_e} d\emph{s}$ is the integral alone the straight trajectory connecting  $k_b$ and $k_e$ points. Moreover, due to the mirror symmetry of this model with $k_y=0$ being the axis of symmetry, one has $|u(k_x,k_y)\rangle=|u(k_x,-k_y)\rangle$, which means $\langle u(k_x,k_y)|\partial_{k_y}|u(k_x,k_y)\rangle=-\langle u(k_x,-k_y)|\partial_{k_y}|u(k_x,-k_y)\rangle$. Therefore, the second term in Eq. (\ref{zak2}) equals to 0, which leads Eq.(\ref{zak2}) to a simple form of
\begin{equation}
\label{zak3}
\phi_{Zak}(\subset)=\oint_S\Omega d\emph{s}.
\end{equation}
Due to the trajectories just encircle a half of the Berry's curvature of a Dirac point in our scheme, the Zak phase $\phi_{Zak}^{\mathrm{\uppercase\expandafter{\romannumeral1}},
\mathrm{\uppercase\expandafter{\romannumeral2}}}
=\varphi^{\mathrm{\uppercase\expandafter{\romannumeral1}},
\mathrm{\uppercase\expandafter{\romannumeral2}}}/2$, which correspond to the geometric phases of the two Dirac points in the first BZ, respectively. Therefore, the CN can be approximated as
\begin{equation}
\label{CN}
C=\frac{1}{\pi}\left(\phi_{Zak}^{\mathrm{\uppercase\expandafter{\romannumeral1}}}
+\phi_{Zak}^{\mathrm{\uppercase\expandafter{\romannumeral2}}}\right).
\end{equation}

\subsection{Control over cold atoms}
For cold atom systems, to create an arbitral interference trajectory in momentum space, we start with a bosonic gas in the state $|\downarrow\rangle$ and bring it into a coherent superposition state $1/\sqrt{2}(|\downarrow\rangle+|\uparrow\rangle)$ using a $\pi/2$ Raman pulse. Then a magnetic field gradient is applied to create a constant force in opposite directions for the two spin components. The moving of atoms in momentum space is described by\cite{L. Tarruell}
\begin{equation}
\label{magnetic field gradient}
\hbar \frac{d k}{d t}= \mp\mu_{\downarrow,\uparrow} \nabla B,
\end{equation}
where $\mu_\uparrow=\mu_\downarrow=\mu$ are the spin magnetic moment and $B$ is the magnetic field. On the other hand, if one properly modulates the lasers, the optical potential is accelerated in the stationary frame. Choosing the optical lattice as the reference frame, the optical potential is stationary and the atoms obtain a inertia force $F$. When a constant force $F$ is produced in the same direction for the two spin components, this result in momentum space is of the form\cite{K. W. Madison}
\begin{equation} \label{accelerating optical lattice}
\hbar \frac{d k}{d t}=-a M,
\end{equation}
where $a$ is the acceleration and $M$ is the mass of the atom. Finally, a combination of the magnetic field gradient and the accelerating optical lattice allows us to arbitrarily control over the evolution of two spin components in momentum space. Experimentally, the characteristic energies of magnetic field gradient and accelerating lattice are at least an order of magnitude smaller than the full bandwidth of optical lattice, hence they don't transfer atoms form lower band to upper band.

\subsection{Detection of the CN}
To detect the CN, a bosonic gas in optical lattice is initially prepared in the Hamiltonian of Eq. (\ref{Haldane model}) with the spin state of $|\downarrow\rangle$  at the bottom of the energy band. Within current technology, one can precisely control a atomic gas to move to anywhere over the BZ \cite{M. Atala}. This detection needs one $\pi$-pluse and two $\pi/2$-pluses, where $\pi$-pulse is equivalent to $\sigma_x$ operation for the atomic spin states $|\uparrow\rangle$ and $|\downarrow\rangle$ and $\pi/2$-pulse is an operation in the form of
\begin{equation}
\label{pi/2}
U_{\pi/2,\phi_{WM}}=\left(   \begin{array}{cc}
1 & i e^{-i \phi_{WM}}\\
i e^{i \phi_{WM}} & 1\\
\end{array}
\right),
\end{equation}
with $\phi_{WM}$ being the controllable phase of the $\pi/2$-pulse.

The detection goes as follows.  We first consider the detection of the contribution from the left Dirac point, as shown in Fig. 1(a). A magnetic field gradient in the $k_x$ direction is applied to create a constant force for the bosonic atoms. Such a constant force leads to a linear evolution of quasimomentum over time, i.e., the Bloch oscillation. Wnen the spin state $|\downarrow\rangle$ moves to the site ($k_x=2\pi/(3\sqrt{3}a), k_y=0$) of the BZ, one uses a $\pi/2$-pulse to bring the state $|\downarrow\rangle$ into a coherent superposition state of $1/\sqrt{2}(|\downarrow\rangle+i |\uparrow\rangle)$. Then, the atoms are moved under the action of the accelerating optical lattice and the gradient magnetic field, the direction of which are $-x$ and $+y$, respectively, and $|aM/(\mu\triangledown B)|=\sqrt{3}$. In this process, the movements of $|\downarrow\rangle$ and $|\uparrow\rangle)$ are exact $-(b_1+b_2)$ and $-b_1$, which leads the superposition state evolves to the sites of ($k_{x,\downarrow}=k_{x,\uparrow}=-4\pi/(3\sqrt{3}a)$, $k_{y,(\downarrow,\uparrow)}=\mp2\pi/(3a)$). The closed trajectory constituted this open trajectory and the straight line between two spin states just encircles a half of Berry's curvature of Dirac point, and thus the superposition states get a Zak phase of $\pm\pi/2$. We proceed to the case of the right Dirac point, as shown in Fig. 1(b). Atoms are now moved to the site ($k_x=-2\pi/(3\sqrt{3}a), k_y=0$). A coherent superposition state of $1/\sqrt{2}(|\downarrow\rangle+i |\uparrow\rangle)$ is created again. After that, under the action of the accelerating optical lattice and the gradient magnetic field with $|aM/(\mu\triangledown B)|=\sqrt{3}$ but the directions of them are $+x$ and $-y$, respectively. In this process, the movements of $|\downarrow\rangle$ and $|\uparrow\rangle)$ are exact $b_1+b_2$ and $b_1$ and the superposition state evolves to the site ($k_{x,(\downarrow,\uparrow)}=4\pi/(3\sqrt{3}a)$, $k_{y,(\downarrow,\uparrow)}=\pm2\pi/(3a)$).

We next link this phase to  a physical observable quantity of atomic spin states in our considered context. For ultracold atoms, bosons in the momentum space are usually considered to be point particles, i.e., the size of which is much smaller than that of the Berry's curvature of a Dirac point, and thus no Zak phase difference between the bosons with same spin states. For this reason, Eq. (\ref{zak2}) for single particle can be directly applied to the case of systems with many bosons as demonstrated in recent experiment \cite{M. Atala}. For the above considered trajectory, the total phase difference between the atomic spin state $|\downarrow\rangle$ and $|\uparrow\rangle$ is $\phi_{tot}=\phi_{Zak}+\phi_{D}+\phi_{Z}$, where $\phi_{D}=\int\triangle E /\hbar d\emph{t}$ is the dynamic phase with $\triangle E$ being the energy difference between two spin states and $\phi_{Z}$ is due to the Zeeman energy of the atoms in an external magnetic field. Due to the symmetry of energy band and the trajectory, the dynamic energy of two spin states are always the same, and thus the dynamic phase is 0. In addition, $\phi_{Z}$ is can be eliminated by a $\pi$-pulse, which flips the spin, in the middle of one straight trajectory while one keeps the direction of the accelerating optical lattice but changes that of the magnetic field direction. Therefore, in this way, the Zak phase is the only contributed component for the total phase. Therefore, the wavefunction has evolved to $\frac{1}{\sqrt{2}}(i |\downarrow\rangle+ e^{i \phi_{Zak}} |\uparrow\rangle)$. To detect the noncyclic Zak phase difference between two spin states, the second $\pi/2$-pulse is applied to the atoms and the atom number in the two spin states $N_{\downarrow,\uparrow}$ are found to be
\begin{equation}
\label{N}
N_{\downarrow,\uparrow}=\frac{1}{2}\left[1\pm\cos(\phi_{Zak}-\phi_{WM})\right].
\end{equation}
When the atoms evolved to the specific site $\mathrm{\uppercase\expandafter{\romannumeral1}}$ and $\mathrm{\uppercase\expandafter{\romannumeral2}}$, the system obtains $\pm \pi/2$ or $\pm 3\pi/2$ noncycle Zak phases, and $N_{\downarrow,\uparrow}=1/2$ without $\phi_{MW}$. The correspondence between the measured atomic population and the Zark phases for the two noncyclic trajectories is listed in table I.

\begin{table}[tb]
\renewcommand{\arraystretch}{1.00}
\caption{The Zak phase $\phi_{Zak}$ and the atom number of the spin up $N_{\uparrow}$ in the different sites for different chern numbers with $\alpha_\pm=[1-\cos(\pi/2\pm\phi_{WM})]/2$. }
\begin{tabular}{|c|c|c|c|c|}
\hline
\multirow{2}{*}{C} & \multicolumn{2}{|c|}{$\mathrm{\uppercase\expandafter{\romannumeral1}}$} &  \multicolumn{2}{|c|}{$\mathrm{\uppercase\expandafter{\romannumeral2}}$}  \\
\cline{2-5}
 & $\phi_{Zak}^{\mathrm{\uppercase\expandafter{\romannumeral1}}}$ & $N_{\uparrow}$ & $\phi_{Zak}^{\mathrm{\uppercase\expandafter{\romannumeral2}}}$ & $N_{\uparrow}$ \\
\hline

\multirow{2}{*}{0} & $\pi/2$ & $\alpha_-$ & $-\pi/2$ & $\alpha_+ $ \\
\cline{2-5}
& $-\pi/2$  & $\alpha_+ $ & $\pi/2$  & $\alpha_- $\\
\hline

$1$ & $\pi/2$ & $\alpha_- $ & $\pi/2$ & $\alpha_-$ \\
\hline

$-1$ & $-\pi/2$ & $\alpha_+ $ & $-\pi/2$ & $\alpha_+ $ \\
\hline
\hline

\end{tabular}
\end{table}

\section{Discussions}
Although the atom numbers $N_{\uparrow}$ without $\phi_{MW}$ in the different sites are the same $1/2$, the detections should display different results in the presence of $\phi_{MW}$. If the geometrical phases of two Dirac points have different signs, which cancel each other out, the total Zak phase is $0$ and $C=0$. In this case, the two Zak phases $\phi_{Zak}^{\mathrm{\uppercase\expandafter{\romannumeral1}}}$ and $\phi_{Zak}^{\mathrm{\uppercase\expandafter{\romannumeral2}}}$ must be obtained as different values.  Therefore, one just need to measure the Zak phase in these special sites for detecting the CN.
In contrast to the case of $C=0$, we next show that the detection display different properties when $C=\pm1$. Because of the same of the geometrical phases of two Dirac points, the atoms should obtain the same Zak phases in these two special sites, that is, if the Zak phase $\phi_{Zak}^{\mathrm{\uppercase\expandafter{\romannumeral1}}}$ is $\pi/2$, $\phi_{Zak}^{\mathrm{\uppercase\expandafter{\romannumeral2}}}$ must be $\pi/2$ for $C=1$. Moreover, the absolute values of Zak phase in the case of  $C=-1$ are the same as that of the $C=1$ case but with opposite sign. In other word, the different functions of $\phi_{MW}$ offer a signal of topologically nontrivial phase characterized by a nonzero CN. Plus or minus of the CN can also be distinguished by comparing the different functions of $\phi_{MW}$ in the different sites. In short, this method can completely distinguish the CN for the considered model.

Experimentally, although one can move atoms to a precise place of the BZ using the current techniques, suppose the atoms do not be moved to the sites ($k_{x,\downarrow}=k_{x,\uparrow}=\pm4\pi/(3\sqrt{3}a)$, $k_{y,\downarrow}=-k_{y,\uparrow}$), the changing trend of $N_{\uparrow}$ with $\phi_{MW}$ is still distinguishable as long as the error of Zak phase is within $\pm\pi/4$, i.e., the $N_{\uparrow}$ without $\phi_{MW}$ is larger than $1/\sqrt{2}$. This means that the sign of Dirac point is also distinguishable and the CN can be obtained as well. Therefore, the requirement of accuracy control over the trajectories is not rigorous. On the other hand, analogues of the method may be applicable in systems with higher CN under the conditions: (1) The Berry curvature distributes in a small region, e.g., within the Dirac cones; and (2)  the geodesic of the needed open trajectories can been easily find out.

\section{Conclusion}
We propose to detect the CN by measuring the atomic population with $\phi_{MW}$ in some special sites of the BZ, which is concerned with the Dirac points. Taking Haldane model for a typical example, we demonstrate that the atomic population as a function of the phase of the final microwave $\pi/2$-pulse and how to distinguish the CN from these functions in the different sites. We further show that this method is experimentally realizable. Therefore, we provide a method to detect topological phases in cold atom context and a newly probable thought to detect the $\mathbb{Z}_2$ topological systems.

\section*{Acknowledgements}

We thank Prof. Shi-Liang Zhu and Dr Zheng-Yuan Xue for many helpful discussions.

\end{document}